\documentclass[10pt,letterpaper]{article}
\usepackage[top=0.85in,left=2.75in,footskip=0.75in]{geometry}

\usepackage{amsmath,amssymb}

\usepackage{changepage}

\usepackage[utf8x]{inputenc}

\usepackage{textcomp,marvosym}

\usepackage{cite}

\usepackage{nameref}
\usepackage[hidelinks]{hyperref}

\usepackage[right]{lineno}

\usepackage{microtype}
\DisableLigatures[f]{encoding = *, family = * }

\usepackage[table]{xcolor}
\usepackage{rotating}

\usepackage{multicol}
\usepackage{multirow}
\usepackage{booktabs}

\definecolor{political}{HTML}{FB6962}
\definecolor{covid}{HTML}{79DE79}
\definecolor{religious}{HTML}{FCFC99}
\definecolor{football}{HTML}{AA9AFC}
\definecolor{generic}{HTML}{A8E4EF}

\usepackage{soul}

\usepackage{array}

\newcolumntype{+}{!{\vrule width 2pt}}

\newlength\savedwidth

\newcommand\thickhline{\noalign{\global\savedwidth\arrayrulewidth\global\arrayrulewidth 2pt}%
\hline
\noalign{\global\arrayrulewidth\savedwidth}}

\raggedright
\usepackage[aboveskip=1pt,labelfont=bf,labelsep=period,justification=raggedright,singlelinecheck=off]{caption}

\makeatletter
\renewcommand{\@biblabel}[1]{\quad#1.}
\makeatother

\usepackage{lastpage,fancyhdr,graphicx}
\usepackage{epstopdf}
\fancyheadoffset[L]{2.25in}
\fancyfootoffset[L]{2.25in}
\begin{document}
\vspace*{0.2in}

\begin{flushleft}
{\Large
\textbf\newline{On the joint effect of culture and discussion topics on X (Twitter) Signed Ego Networks} 
}
\newline
\\
Jack Tacchi\textsuperscript{1,2},
Chiara Boldrini\textsuperscript{1},
Andrea Passarella\textsuperscript{1},
Marco Conti\textsuperscript{1}
\\
\bigskip
\textbf{1} Institute of Informatics and Telematics, Consiglio Nazionale delle Ricerche, Pisa, Italy
\\
\textbf{2} Scuola Normale Superiore, Pisa, Italy
\\
\bigskip

%
%





* jacktacchi@gmail.com

\end{flushleft}
\section*{Abstract}
  Humans are known to structure social relationships according to certain patterns, such as those described by the Ego Network Model (ENM). These patterns result from our innate cognitive limits and can therefore be observed in the vast majority of large human social groups. Until recently, the main focus of research was the structural characteristics of this model. The main aim of this paper is to complement previous findings with systematic and data-driven analyses on the positive and negative sentiments of social relationships, across different cultures, communities and topics of discussion. A total of 26 datasets were collected from the X social media platform for this work. It was found that, contrary to previous findings, the influence of culture is not easily ``overwhelmed'' by that of the topic of discussion. However, more specific and polarising topics do lead to noticeable increases in negativity across all cultures. These negativities also appear to be stable across the different levels of the ENM, which contradicts previous hypotheses. Finally, the number of polarising topics amongst those most discussed seems to be a good predictor of the overall negativity of users' relationships, although this finding is somewhat limited by the small sample size of the data.


\section{Introduction}
\label{sec:introduction}

Throughout human history, our ability to communicate has been a defining trait of our species. 
Communication has an immeasurable impact on our behaviour and its importance is evident at all levels of society.
It affects everyday interactions between individuals as well as how our societies are organised. 
Very importantly, it is primarily through communication and interaction that we build our social networks.
Given the omnipresence of communications, one might expect a great deal of variety in the resulting patterns of social relationships. 
However, some intriguingly common patterns can be observed, resulting from our innate cognitive limits.

One such pattern can be observed between the proportional size of a species' neocortex and the size of a social group that that species can maintain~\cite{Dunbar_1992}.
The link between these two variables is known as the Social Brain Hypothesis and it further posits that if a social group grows beyond its maintainable size, it will inevitably start to break down into smaller, less cognitively demanding collectives.
This phenomenon is so ingrained in our evolutionary neurology that it has been observed not just for humans but also for many other types of primates, and even some species of birds~\cite{Dunbar_1998}.
By extrapolating the observed group sizes of animals up to the size expected for an animal with the neocortex size of a human, one would expect our own social group size limit to be around 150 (known as Dunbar's number).
Indeed, 150 is a common unit size across human social structures and has been recorded in contexts as diverse as modern-day militaries' company sizes and traditional hunter-gatherer communities from 5 different continents~\cite{Dunbar_1993}.
More recently, this has also been observed in online social networks~\cite{West_2020}.

\begin{figure}[h]
  \centering
  \includegraphics[scale=0.25]{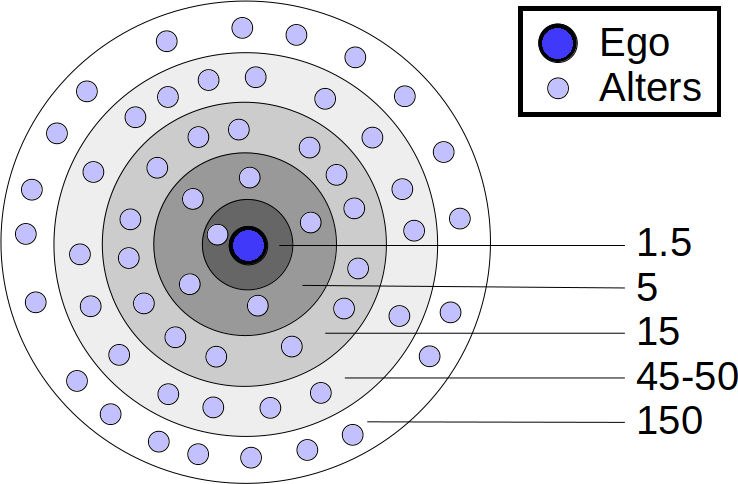}
  \vspace{1em}
  \caption{A graphical representation of the Ego Network Model. The dark blue dot at the centre represents the Ego, it is surrounded by many smaller light blue dots, representing the Alters. These are organised into the circles of the ENM, and labelled with their expected numbers.}
  \label{fig:ego_network_model}
\end{figure}

Additionally, when a human social network is viewed from the point of view of a single individual, Dunbar's number once again emerges.
Indeed, the number of annually active relationships being maintained is seldom far from 150.
If these relationships are then organised by the strength of their connection to the initial subject, a series of concentric circles of increasing size but decreasing connection strength will almost inevitably be observed~\cite{Dunbar_1995}.
This individual-focused representation of a social network is called the Ego Network Model (ENM), with the individual in question being referred to as the Ego (from which the model takes its name) and their connections being called Alters.
What's more, because these circles are another emergent pattern resulting from neocortical limits, their sizes are incredibly regular. 
For humans, these sizes are 5 (support clique), 15 (sympathy group), 45-50 (affinity group) and 150 (active network), with each subsequent circle size increasing by a ratio of around 3. 
Ego networks have also been studied for Online Social Networks. 
Despite the ease of establishing social connections, data-driven studies have shown that the ENM can be found also in Online Social Networks~\cite{Dunbar_2015}. 
Further, Ego Networks feature the same layered structure, and the sizes of the layers align with what has been found in ``offline'' social networks. 
The only notable difference is the emergence of an additional inner-most layer, of average size 1.5~\cite{Dunbar_2015}. 
Similarly, Ego Networks have also been observed in networks of phone calls and text messages~\cite{Heydari_2024}.
As previously observed~\cite {Miritello_2013}, this confirms that Ego Network structures are determined by human cognitive limits no matter how advanced the ``tools'' used to facilitate social interactions.
Fig~\ref{fig:ego_network_model} shows a standard representation of an Ego Network from an online context.

Such egocentric representations differ from sociocentric alternatives in that they focus on the networks from an individual's point of view, as opposed to taking a wider view of the network and all of its interconnected users as a whole.
So, while sociocentric approaches primarily focus on the structure of a network, egocentric approaches follow the tenet that each individual in a network is at the centre of their own personal community~\cite{Perry_2018}.
Egocentric methods therefore focus principally on individual contexts within a network.
What's more, the Ego Network Model is a specific model within egocentric approaches that differs from other works, such as~\cite{Perry_2018} and~\cite{McAuley_2014}, in that it does not consider the connections between Alters, even if they exist in reality.
This can provide an advantageous perspective when observing many phenomena in social networks.
Indeed, Ego Networks have been used alongside sociocentric representations to gain additional insights that would otherwise be much harder to obtain, especially in situations where there are important central nodes~\cite{Socievole_2012}.
An example of this can be found in that certain aspects of an individual's Ego Networks have been used to predict core personality traits.
Such as a high turnover of Alters correlating to individuals with high degrees of openness~\cite{Centellegher_2017}.
So Ego Networks are clearly a useful tool for enabling further discoveries, and the more we improve our understanding of this tool, the better we will be able to use it.

Of course, determining how to measure the strength of the connection between an Ego and their Alters is of key importance to the ENM.
What's more, it is not something that can be objectively defined.
Fortunately, there is a renowned definition that has become the standard for research on Ego Networks: the tie strength between two individuals is the (probably linear) combination of the time spent maintaining the relationship, its emotional intensity, its level of intimacy and the reciprocal services it generates~\cite{Granovetter_1973}. 
Given the obvious difficulty of objectively measuring the tie strength between two individuals along the dimensions defined in~\cite{Granovetter_1973}, traditionally, the Ego-Alter contact frequency has often been used.
While it has been found that this is a suitable proxy metric~\cite{Gilbert_2009} for relationship strength, it may miss a lot of potentially important information.
One could reasonably expect contact frequency to be strongly correlated with the time spent on the relationship (perhaps explaining why it is such a good proxy metric) but the qualitative aspects of relationships might be completely ignored.

To address this problem, some recent works~\cite{Tacchi_2022, Tacchi_2023} have taken a bottom-up approach by applying sentiment analysis models to generate a label for each interaction within a relationship and then, using a psychology-based threshold~\cite{Gottman_1993}, inferring a \emph{sign} for the relationship as a whole.
This threshold, referred to here as the ``Golden Interaction Ratio'', comes from observations that relationships with more than 1 negative interaction for every 5 positive interactions (roughly 17\%) tend to cause myriad problems for those involved.
For marriages, this means a significant increase in the probability of divorce~\cite{Gottman_1993}, and for parental relationships, can lead to the child developing behavioural problems and/or struggling in school~\cite{Hart_1995}.
Subsequent research has found that this method could be biased if there is an imbalance in power between the individuals involved~\cite{Gottman_1999}, however, this is unlikely to be the case for the majority of the online relationships used in this work.

Using the aforementioned inferred signs, some novel insights were able to be obtained, providing solid grounds for the definition of a Signed Ego Network Model (hereafter referred to as SENM).
For instance, despite the consistency of the presence of the ENM, the distribution of the signs of its relationships can vary dramatically.
Indeed, one past paper made a concentrated effort to compare levels of negativity across multiple different cultures and communities~\cite{Tacchi_2023}. 
The networks' negativities were observed to vary between 49.90\% and 69.47\%.
What's more, it was found that, although there were some differences due to culture, these appeared to be overwhelmed by the impact of the topic around which a community was formed (e.g. journalism or reality TV). 
Furthermore, the influence of the topic was stronger the more negative it was, with the differences between cultures decreasing dramatically for the most negative topics.
However, in the absence of a specific topic-based community, cultural differences were still very much observable.
A concept that runs parallel to topic's negativity is how polarising it is.
Here, polarisation can be defined as the tendency of group members with opposite viewpoints to become more and more ideologically separated as they engage with a given topic~\cite{Sunstein_1999}.

Understanding these cultural differences becomes even more important the more different cultures interact and, in the globally connected modern world, interculture communications occur more frequently than ever.
Although, given that the SENM is a very recent development, investigations between culture, and other types of communities, have so far been very preliminary.
In addition to cultural differences, and in contradiction with previous expectations, the most negative circles of the SENM tend to be the innermost ones~\cite{Tacchi_2022}.
In the conventional ENM, the inner circles are where our most important and trusted connections lie~\cite{Sutcliffe_2015}, so this was an unexpected finding.
As with the negativity of each network as a whole, a better understanding of how the negativities of the different circles within the SENM change between cultures may reveal non-trivial and impactful insights.

This paper is an extension of the aforementioned cultural analysis of the SENM~\cite{Tacchi_2023}, which aims to enhance many of the previous findings with regard to differences arising from culture and topic.
Specifically, the previous paper conducted some basic analyses about the most popular words and hashtags used by different users while this extension further leverages a deep learning model, BERTopic, to provide a more rigorous analysis of numerous combinations of topic and culture.
This is also a more nuanced approach compared to the previous study of the most popular words and hashtags.
Fifteen additional datasets were collected with the specific purpose of gaining a better understanding of certain phenomena that were observed in the initial paper.
This has resulted in 4 main novel discoveries. The first of these is that, although there are some differences in negativity caused by both culture and topic, cultural differences do not seem to get ``overwhelmed'' by topics as easily as previously thought. 
Instead, the effects of culture on the negativity of a SENM appear noticeable the majority of the time and are only overpowered by very polarising topics. 
Next, when a network is centred around a more specific or polarising topic, it invariably displays an increase in negativity at all levels of the SENM; this effect is observable for all cultures. 
Thirdly, the level of negativity across the circles of the SENM does not appear to be less stable for more negative or specific topics, as was found previously. 
Finally, the number of polarising topics within a dataset's top 20 most popular topics can be used to predict how negative the relationships in it are overall, regardless of how negative the corresponding individual tweets are.

\section{Methodology}
\label{sec:methods}

\subsection{Signing Relationships}
\label{sec:signing_relationships}

To determine the sign of each relationship, a bottom-up approach, echoing the methods of previous SENM research~\cite{Tacchi_2022, Tacchi_2023}, was employed. 
First, this method groups all interactions between each Ego and Alter. 
Then, it leverages the well-established field of sentiment analysis to provide a sentiment (``positive'', ``neutral'' or ``negative'') based on the text of each interaction individually.

While the exact choice of model used for this step has varied in previous work, it has been shown that the relationship signs are not overly affected by this choice~\cite{Tacchi_2024}. Given the use of multilingual data in this paper, a polyglot model was used: XLM-T~\cite{Barbieri_2022}. 
This model is based on another multilingual model, XLM-R~\cite{Conneau_2019}, which was trained on Wikipedia texts from 100 different languages. XLM-T was then hyper-tuned for dealing with Tweets by further training it on 198 million Tweets from over 60 languages. 
Using the XLM-T, a list of sentiments for each interaction in every Ego-Alter pair was computed.

The final step is to apply the Golden Interaction Ratio of 17\% (mentioned in Section~\ref{sec:introduction}). This results in a positive sign for relationships with percentages of negative interactions below or equal to this threshold, and a negative sign for those above it.

\subsection{Computing Signed Ego Networks}

To generate a Signed Ego Network, the unsigned version of it first needs to be obtained. As described in Section~\ref{sec:introduction}, this is done by organising Alters around their corresponding Ego, based on their tie strength. In purely practical terms, this means clustering the Alters based on the Ego's contact frequency with them. Multiple algorithms have been used for clustering the Alters in Ego Network research. The one used in the current work, the MeanShift algorithm~\cite{Fukunaga_1975}, is one of the most commonly used~\cite{Arnaboldi_2015,Toprak_2022,Tacchi_2024} and also has the added benefit of automatically determining the optimum number of circles (clusters) for an Ego.

After computing both the unsigned Ego Networks and the relationship signs (obtained via the steps outlined in Subsection~\ref{sec:signing_relationships}), it is a simple matter of matching the relationship signs with the corresponding Ego-Alter pairs in each Ego Network in order to obtain the Signed Ego Networks.

\subsection{Topic Analysis}
\label{sec:method_topic_analysis}

Previous work has found that the percentage of negative relationships in a dataset can be dependent on the type of community that the dataset contains. For instance, communities based around more specific topics, such as journalism or reality TV, will likely be more negative than those of a more arbitrary or generic nature, such as geographical region~\cite{Tacchi_2023}. Thus, an in-depth analysis of the topics within some of the larger generic datasets used in this paper was conducted (see Section~\ref{sec:datasets}), to see whether this effect can be replicated by focusing on the subcommunities within a generic dataset.

In order to accurately detect the main topics of discussions within multilingual data in a standardised way, the BERTopic model~\cite{Grootendorst_2022} was employed. 
BERTopic is a topic modelling tool that uses a mixture of transformers and TF-IDF to identify important topics and terms within a collection of natural-language documents.
It then uses HDBSCAN, a hierarchical clustering algorithm~\cite{Ester_1996}, to cluster the identified topics together based on similarity. 
Specifically, if two topics are separated by a distance that is less than a predefined distance (referred to here as the clustering epsilon) then they are merged together.
BERTopic was compared against 5 other state-of-the-art topic models in terms of two staple metrics: Topic Coherence~\cite{Lau_2014} and Topic Diversity~\cite{Dieng_2020}. 
These measure how well a model's grouped terms fit with one another and how much variety there is among grouped words. 
BERTopic was found to consistently outperform the other models for Topic Coherence while also remaining very competitive for Topic Diversity~\cite{Grootendorst_2022}.
In order to check BERTopic's performance for the data used in this paper, both Coherence and Diversity are measured in Subsection~\ref{sec:coherence_and_diversity}.

To optimise BERTopic for multilingual data, the default transformer model it uses can be replaced with paraphrase-multilingual-MiniLM-L12-v2~\cite{Reimers_2019}. 
This is a sentence-transformer model that is able to accurately process data in over 50 languages~\cite{Reimers_2020}.
Unfortunately, paraphrase-multilingual-MiniLM-L12-v2 can only take in the first 384 tokens from each document, making it impossible to pass in an entire User Tweet Timeline (which often contains a few thousand Tweets, each of up to 280 characters) as a single document.
Tweets were therefore parsed individually, meaning that each Tweet was treated as being entirely distinct from all the others, even if they were created by the same user. 
An unfortunate consequence of this is that some of the topics may appear to be undeservedly important if, for example, they are being spam-tweeted by an individual user.
In an effort to mitigate this, and in order to get a perspective of the topics being mentioned by all users in each dataset, the top 200 topics provided by BERTopic were collected and the 20 topics with the largest numbers of unique users were then chosen as the focus of analysis using the 3 large Generic Users datasets (see Subsection~\ref{sec:topic_analysis}). 
Effectively, this meant that each topic was counted a maximum of once per user, rather than once per Tweet related to the topic.  These topics are listed and discussed in Subsection~\ref{sec:topic_analysis}.
Indeed, given the nature of the data, it is unsurprising that the number of both Tweets and users follows Zipf's law~\cite{Zipf_1936}, with a lot of Tweets and users engaging in a handful of important topics and a long tail of topics with only a few users and Tweets.

The models, user topics and edgelists used for the topic analyses can be found at \url{https://zenodo.org/records/14336102}.

\section{Datasets}
\label{sec:datasets}

\subsection{Data Source}
\label{sec:data_source}

Every dataset included in this paper was collected from the X social media platform (formerly known as Twitter). 
The collection and analysis method complied with the terms and conditions of the source of data.
Historically, this platform has been a consistent source of Ego Network data for over a decade~\cite{Arnaboldi_2013, Toprak_2022} and used to allow quick and easy access to a huge amount of public social communications data from users all across the globe via its API. 
In addition to this, the site allows its users to interact with one another in a variety of ways that make the data especially suitable for Ego Network research. 
Namely, these are Replies, Mentions and Retweet, which allow specific users to be tagged whenever a communication is made, in turn allowing Egos and Alters to be easily mapped and tracked to individual interactions.

All of the data used in this paper were collected using the, now defunct, free academic version of the Twitter API that, among others, provided two important endpoints. 
These were Twitter Search, which took in a search query and provided a stream of Tweets relating to it (in reverse chronological order), and User Tweet Timelines, which took in the ID or name of a specific user and returned the entirety of their publicly created Tweets. 
Of course, this latter also includes Tweets that contain no communication data between users, which is useless for the current research.
Fortunately, X makes it possible to know when users are specifically contacting another user or users, via the use of ``Mentions'', ``Replies'' and ``Retweets''. 
Although the last of these three options is only considered a communication for this work if the authoring user also adds some text (known as a ``Quote Retweet''). 
This is to ensure there is at least some cognitive involvement from the Ego and is a standard approach in the related literature~\cite{Tacchi_2022}.
Further, it is important to note that this version of the API, unlike its predecessors, had no upper limit on the number of Tweets returned within a User Timeline.
This means that, for each user that was collected, every Tweet they ever published was collected (unless any Tweets were deleted or restricted prior to collection).

The datasets have been split between those that existed prior to this paper and those that were collected as part of it. 
They are described in further detail, in Subsections~\ref{sec:datasets_old} and~\ref{sec:datasets_novel} respectively, and some descriptive statistics are provided in Table~\ref{tab:descriptives_active_network} (after all preprocessing steps mentioned in Subsection~\ref{sec:preprocessing}).

\begin{table}
  \caption{Number of Egos, Relationships and Interactions in the active Ego Networks, after all preprocessing steps.}
  \label{tab:descriptives_active_network}
  \begin{tabular}{|l|l|r|r|r|}
    \hline
    \textbf{Dataset Type} & \textbf{Region} & \textbf{Egos} & \textbf{Relationships} & \textbf{Interactions} \\
    \thickhline
    \multirow{1}{*}{Baseline} & -- & 4,049 & 574,585 & 8,593,290\\
    \hline
    \multirow{4}{*}{Geographical} & Mediterranean & 878 & 120,068 & 2,191,666\\
     & South America & 217 & 25,205 & 441,158\\
     & Northern Europe & 552 & 82,237 & 1,273,881\\
     & West Africa & 396 & 55,884 & 884,321\\
    \hline
    \multirow{3}{*}{Reality TV} & Italy & 160 & 18,884 & 291,213\\
     & Brazil & 154 & 15,685 & 234,734\\
     & Netherlands & 230 & 24,082 & 441,694\\
    \hline
    \multirow{3}{*}{Journalists} & Italy & 203 & 30,409 & 489,008\\
     & Brazil & 154 & 20,348 & 278,631\\
     & Netherlands & 1,316 & 179,668 & 2,702,275\\
    \hline
    \multirow{3}{*}{Generic Users} & Italy & 2,740 & 266,701 & 2,133,608\\
    & Brazil & 8,223 & 820,165 & 6,561,320\\
    & Netherlands & 9,278 & 863,187 & 6,905,496\\
    \hline
    \multirow{4}{*}{Weather} & Italy & 518 & 42,168 & 337,344\\
    & Brazil & 598 & 42,553 & 340,424\\
    & Netherlands & 255 & 14,427 & 115,416\\
    & Nigeria & 363 & 22,628 & 181,024\\
    \hline
    \multirow{4}{*}{Football} & Italy & 1,320 & 141,565 & 1,132,520\\
    & Brazil & 1,024 & 117,199 & 937,592\\
    & Netherlands & 1,910 & 156,102 & 1,248,816\\
    & Nigeria & 159 & 14,206 & 113,648\\
    \hline
    \multirow{4}{*}{Politics} & Italy & 2,004 & 218,005 & 1,744,040\\
    & Brazil & 482 & 44,333 & 354,664\\
    & Netherlands & 1,256 & 151,922 & 1,215,376\\
    & Nigeria & 866 & 65,846 & 526,768\\
    \hline
  \end{tabular}
\end{table}


\subsection{Pre-existing Datasets}
\label{sec:datasets_old}

As a starting point for the current work, 11 datasets were collected from previous Ego Network research. These represent data from a mixture of different countries. Additionally, they can be easily sorted between ``generic'' and ``specialised'' users. Generic users are those who use the social media platform (in this case X) for predominantly social reasons, whereas specialised users use it for professional reasons. These two types of users have been shown to exhibit differing behaviours in online contexts~\cite{Toprak_2022} and it is therefore important to bear this in mind during the collection and analysis of the data.

\subsubsection{Baseline}
The first of the datasets used in this paper is referred to in this paper as Baseline because its initial purpose was to obtain a baseline measurement of an X user's percentage of negative relationships~\cite{Tacchi_2022}. 
It was collected using a snowball sampling methodology with an initial set of 31 seed users. The User Tweet Timelines of these users were then gathered, followed by those of their Alters, then of their Alters' Alters, and so on, until the collection period ended. The seed users were randomly selected from another large Ego Network dataset, which was itself a snowball sampling that used Barack Obama's official X account as its seed user~\cite{Arnaboldi_2013}. The Baseline dataset is composed of the User Tweet Timelines of the collected users\footnote{All Tweet IDs from this dataset are provided at \url{https://zenodo.org/records/7717006}.} and was collected between the 27\textsuperscript{th} April and the 25\textsuperscript{th} May 2022.
Recalling from Section~\ref{sec:data_source}, although the Timelines were collected between these dates, they would include the vast majority of each Tweet that the users ever made.
Thus, they often spanned multiple years, and this is true of all the datasets used for this paper.

\subsubsection{Geographical}
The next 4 datasets were collected using the same snowball collection methodology as the Baseline dataset~\cite{Tacchi_2023}. However, rather than obtaining a baseline of the entirety of X, the purpose of these datasets was to gauge the negative relationships of specific regions. This was achieved by using seed users from countries in 4 geographically and culturally distinct regions: Mediterranean (Spain, France, Italy, Greece), South America (Brazil, Colombia, Venezuela), Northern Europe (Germany, Netherlands, Sweden) and West Africa (Nigeria, Senegal, Ghana)\footnote{All Tweet IDs from these datasets are provided at \url{https://zenodo.org/records/7717047}.}. These datasets were collected between the 16\textsuperscript{th} June 2022 and the 26\textsuperscript{th} July 2022.

\subsubsection{Reality TV} 
The next set of datasets is the first that targets a specific type of user, in this case, those who follow reality TV shows. 
This was done by using Twitter Search to query for hashtags related to shows that were popular in 3 target countries: Italy (\#XF2022, \#GFVIP), Brazil (\#XFactorBR, \#BBB22) and the Netherlands (\#HollandsGotTalent, \#IkVertrek)\footnote{All Tweet IDs from these datasets are provided at \url{https://zenodo.org/records/7716860}.}~\cite{Tacchi_2023}. 
The previous work did not include a dataset corresponding to the West Africa region. Seed users were manually identified for each dataset using the results of the aforementioned search and User Tweet Timelines were then collected using the same snowball sampling methodology as the Baseline. The Reality TV datasets were collected between the 21\textsuperscript{st} and the 29\textsuperscript{th} January 2023.

\subsubsection{Journalists}
The final set of pre-existing datasets used in this paper is taken from a paper that investigated the differences between generic users and journalists on X~\cite{Toprak_2022}. 
Unlike the snowball sampling method of the previous datasets, the users in this set were obtained via lists of verified journalist X accounts. 
The Journalist datasets contain the User Tweet Timelines of journalists from Italy, Brazil and the Netherlands respectively. 
Again, there was no corresponding dataset available for West Africa. 
These datasets were collected between the 14\textsuperscript{th} and the 17\textsuperscript{th} January 2018

\subsection{Novel Datasets}
\label{sec:datasets_novel}

To complement the pre-existing datasets, an additional 15 datasets were collected\footnote{All Tweet IDs from these datasets are provided at \url{https://zenodo.org/records/10605838}.}. 
These were chosen to better understand phenomena observed in previous work~\cite{Tacchi_2023}. 
Specifically, they were selected to compare the differences in negativity arising between topics of differing levels of controversy, and their interplay with cultural factors.
One important caveat to this is that, in order to facilitate the analysis of these datasets, the assumption is made that they are all independent, i.e. non-overlapping.
While this is a common research assumption, it has been pointed out that this can be somewhat naive when dealing with real-world data~\cite{Lancichinetti_2009}.
This assumption is discussed again in Subsection~\ref{sec:coherence_and_diversity}.

All of these datasets used the same snowball sampling collection method as the Baseline dataset. 
All randomly selected seed users were manually checked for suitability and replaced if they were thought to be spammers, bots, businesses or individuals from outside the desired country.

\subsubsection{Generic Users}
The first set of novel datasets was created to fulfil a similar rationale to that of the pre-existing Geographical datasets but, instead of collecting users from multiple countries within a large region, these datasets took a more precise focus and only collected users from specific countries. 
Specifically, we picked one country for each of the macro-areas in the Geographical dataset, focusing specifically on Italy, The Netherlands, Brazil and Nigeria. These countries were chosen by counting the number of active X users in the Baseline dataset for each country and selecting the highest for each region. 
These datasets were collected between the 3\textsuperscript{rd} May and the 4\textsuperscript{th} June 2023. Unfortunately, the academic version of the Twitter API was disabled around the end of this period, which prevented the collection of a Nigerian Generic Users dataset.

Because it is not possible to manually check the country of every user in the datasets and, as the snowball sampling method allows for the possibility of users to be collected from outside the target country, some tests were performed after the collection of the Generic User datasets. 
These involved checking the self-declared location and X-detected main language of each user. 
For the Brazilian and Dutch datasets, the vast majority of users' locations and languages were as expected. However, the Italian dataset showed a significant proportion (around a third) of users from the UK and USA. To combat this, users were removed from this dataset if their location contained ``UK'', ``London'', ``USA'', ``DC'' or ``CA'' or if their main language was English.
Unfortunately, this resulted in a much smaller dataset than the other two Generic Users, however, this is still one of the larger datasets used in this paper. 
Indeed, these datasets specifically included much larger amounts of data so that the subcommunities of these datasets could be examined (see Subsection \ref{sec:topic_analysis}).

\subsubsection{Weather}
The next set contains slightly less generic datasets focused around the weather. The initial seed users were randomly selected from users who commented on the posts of local weather forecasting X accounts. 
These accounts were Meteo Italia (@meteo\_italia7) and meteo.it (@wwwmeteoit) for Italy, MetSul Meterologia (@metsul) for Brazil and Weer \& Radar Nederland (@weerenradar\_nl) for the Netherlands. 
For Nigeria, it was not possible to find a weather-related account that generated more than a few comments from other users, so the seed users were collected using a Twitter Search for ``weather nigeria'' instead. 
These datasets were collected between the 21\textsuperscript{st} and 25\textsuperscript{th} April 2023.

\subsubsection{Football}
The first of the two specialised sets of novel datasets is themed on Football. 
For these datasets, seed users were collected from users commenting on posts made by popular local football teams: Juventus (@juventusfc) for Italy, Regatas do Flamengo (@Flamengo) for Brazil, Ajax (@AFCAjax) for the Netherlands and Enyimba (@EnyimbaFC) and Plateau United (@plateau\_united) for Nigeria. 
Nigeria was the only country whose most popular football teams were from foreign countries, such as Manchester United, Chelsea and Barcelona. 
This was also why a second team was included, as it was not possible to get enough seed users using a single Nigerian football team. 
These datasets were collected between the 20\textsuperscript{th} April and the 22\textsuperscript{nd} May 2023.

\subsubsection{Politics}
The final set of datasets used in this paper focuses on Politics. Seed users for these were taken from users commenting on posts made by political parties in each of the target countries. 
In order to get a broader image of the general political discussions of each country, rather than that of any single political party, a list of seed users was generated for multiple parties for each country, and the final seeds used were selected randomly from these lists, with a minimum of 5 users and a maximum of 12 users from each party's list.
For Italy, the chosen parties were Fratelli d'Italia (@FratellidItalia), Lega Salvini (@LegaSalvini) and Partito Democratico (@pdnetwork), for Brazil they were Partido Liberal (@PartidoLiberal), Movimento Democrático Brasileiro (@MDB\_Nacional) and Partido dos Trabalhadores (@ptbrasil), for the Netherlands Volkspartij voor Vrijheid en Democratie (VVD), Democraten 66 (@D66), Christen-Democratisch Appèl (@cdavandaag) and for Nigeria they were All Progressives Congress (@OfficialAPCNg), Peoples Democratic Party (@OfficialPDPNig), Labour Party (@OfficialPDPNig) and New Nigeria Peoples Party (@OfficialNNPPng). 
These datasets were collected between the 27\textsuperscript{th} April and the 20\textsuperscript{th} May 2023.

\subsection{Preprocessing}
\label{sec:preprocessing}

\subsubsection{Non-Human Users}
\label{sec:non-human_users}
Unfortunately, not all accounts on X are controlled by human individuals. When sampling such large quantities of accounts, many spammers, bots and groups (such as businesses) will inevitably end up among them.
Obviously, such undesired users do not have the same cognitive constraints as a single human, meaning that they will not display the ENM structure within their communications.
Therefore, it is an important preprocessing step to identify and remove any non-human accounts.
Given the common usage of X for the collection of Ego Network data, this is a common problem. 
A standard method~\cite{Cortes_1995} of filtering out non-humans is to use a Support Vector Machine (SVM), trained on a set of 500 users, to label each account as ``people'' or ``other''. 
Both this filtration method and the training set have been established in previous ENM research~\cite{Arnaboldi_2013}\footnote{A github repository containing the data used to train the SVM is available at the following link: \href{https://github.com/valearna/egonetworks/tree/master/datasets}{github.com/valearna/egonetworks/tree/master/datasets}}.
The SVM was trained using 115 features for each user: 15 profiles features, such as their total number of tweets and the age of their account, and 100 timelines features, such as the mean length of their tweets and the percentage of their tweets that are mentions, replies and retweets.
Using k-fold cross-validation (with a k value of 5), this model originally achieved an accuracy of 81.3\%.
Once the SVM has been trained, it can be used on the collected data to obtain ``other'' accounts and remove them.

This preprocessing step was performed for all datasets except the Journalists. This is because those users were collected from a verified list of known accounts of journalists and therefore did not contain any undesired types of users.

\subsubsection{Irregular Egos}
The second preprocessing step was to remove irregular users.
Egos who spend little time on X or who engage with it infrequently will not have a fully formed Ego Network on the platform, which is problematic for the analyses of this paper. Therefore, Egos were removed if the total length of their User Tweet Timeline was less than 2,000 Tweets, if their Timeline spanned fewer than 6 months, or if they tweeted less than once every 3 days for more than half of the months they were active. These parameters have previously been shown to be appropriate for preparing data for Ego Network analysis~\cite{Arnaboldi_2015}.

Unlike the non-human filter, inactive users were removed from all the datasets including the journalists. This is because, although the list of journalists was verified, it is still possible that some of those users were not fully engaged with the platform when their data was collected.

\subsubsection{Inactive Relationships}
As stated in Section~\ref{sec:introduction}, an individual's Ego Network is expected to contain up to around 150 Alters.
However, this is not the number of users that an individual will interact with throughout their lifetime. 
Humans will interact with a significantly larger number of Alters but many of these will be one-off instances that will have little to no effect on cognitive load.
This threshold has traditionally been defined as once per year~\cite{Hill_2003} and, as it results from our innate cognitive constraints, should not be expected to vary significantly between different cultures or groups.
Thus, Alters were removed if their Ego interacted with them, on average, less than once annually.
Again, this preprocessing step was carried out for all of the datasets used in this paper.

\section{Results}
\label{sec:results}

\subsection{Unsigned Ego Network Analysis}
First of all, the unsigned structures of the Ego Networks in each dataset were analysed, to ensure that there are no anomalies in the datasets. 
The first step in this process is to calculate the mean active network sizes and mean numbers of circles.
As can be seen in Table~\ref{tab:optimum_circle_and_mean_network_size}, the mean active network sizes of the datasets mostly fall around the 90 to 120 range.
Although this is lower than the 150 that would be expected according to the standard ENM, it is common to observe slightly smaller active networks when using social media data~\cite{Toprak_2022,Tacchi_2023}.
This is because a user's true active network will inevitably include real-world relationships that are not present online.
Similarly, the mean number of circles is close to 5 for all the datasets, which is the expected number for online data~\cite{Dunbar_2015}.
In fact, 5 is the closest integer for all of the datasets except the Italian Journalist and the Dutch and Nigerian Weather datasets, for which it is 6, 4 and 4 respectively.
Slight variations around 5 are commonly found in the literature~\cite{Arnaboldi_2013, Dunbar_2015} and can be attributed to psychological and behavioural differences between individuals (for example, how social they are and, for social media datasets, how engaged they are with the given platform).

\begin{table}
\begin{adjustwidth}{-1.25in}{0in}
  \caption{Mean active network sizes, number of optimum circles [95\% confidence intervals] and number of Egos with 5 circles}
  \label{tab:optimum_circle_and_mean_network_size}
  \begin{tabular}{|l|l|l|l|r|}
    \hline
    \textbf{Dataset Type} & \textbf{Region} & \textbf{Mean Network Size} & \textbf{Mean \# Circles} & \textbf{\# 5-Circle Egos}\\
    \thickhline
    \multirow{1}{*}{Baseline} & -- & 99.05 [96.49, 101.60] & 4.81 [4.78, 4.84] & 1,160\\
    \hline
    \multirow{4}{*}{Geographical} & Mediterranean & 109.68 [103.78, 115.58] & 5.11 [5.03, 5.18] & 374\\
    & Northern Europe & 119.60 [112.68, 126.51] & 5.10 [5.01, 5.19] & 275\\
    & West Africa & 102.32 [94.94, 109.69] & 5.00 [4.90, 5.10] & 206\\
    & South America & 101.94 [92.89, 110.99] & 4.85 [4.73, 4.96] & 130\\
    \hline
    \multirow{3}{*}{Reality TV} & Italy & 103.65 [88.05, 119.25] & 5.48 [5.25, 5.71] & 40\\
    & Brazil & 96.63 [85.11, 108.14] & 5.42 [5.22, 5.62] & 48\\
    & Netherlands & 98.63 [86.41, 110.85] & 5.20 [5.01, 5.39] & 62\\
    \hline
    \multirow{3}{*}{Journalist} & Italy & 120.10 [110.49, 129.70] & 5.72 [5.20, 5.93] & 51\\
    & Brazil & 116.48 [103.95, 129.01] & 5.46 [5.27, 5.65] & 50\\
    & Netherlands & 122.69 [118.42, 126.96] & 5.45 [5.42, 5.51] & 440\\
    \hline
    \multirow{3}{*}{Generic Users} & Italy & 101.48 [97.36, 105.60] & 4.88 [4.83, 4.94] & 699\\
    & Brazil & 101.86 [99.60, 104.11] & 4.96 [4.93, 4.99] & 2,208\\
    & Netherlands & 102.64 [100.12, 105.16] & 4.79 [4.76, 4.82] & 2,274\\
    \hline
    \multirow{4}{*}{Weather} & Italy & 102.86 [91.11, 114.62] & 4.67 [4.55, 4.78] & 106\\
    & Brazil & 90.17 [81.54, 98.80] & 4.54 [4.45, 4.78] & 142\\
    & Netherlands & 88.06 [73.50, 102.63] & 4.28 [4.13, 4.43] & 56\\
    & Nigeria & 81.60 [70.89, 92.31] & 4.36 [4.24, 4.49] & 68\\
    \hline
    \multirow{4}{*}{Football} & Italy & 107.44 [101.74, 113.13] & 5.08 [5.01, 5.15] & 344\\
    & Brazil & 107.01 [101.74, 113.13] & 5.27 [5.20, 5.35] & 281\\
    & Netherlands & 88.65 [83.82, 93.48] & 4.66 [4.60, 4.72] & 457\\
    & Nigeria & 110.76 [96.58, 124.93] & 4.80 [4.59, 5.01] & 41\\
    \hline
    \multirow{4}{*}{Politics} & Italy & 104.00 [99.49, 108.51] & 5.12 [5.06, 5.18] & 510\\
    & Brazil & 100.88 [92.76, 109.01] & 4.91 [4.80, 5.03] & 134\\
    & Netherlands & 111.31 [104.63, 118.00] & 5.24 [5.16, 5.32] & 332\\
    & Nigeria & 87.55 [82.27, 92.83] & 4.67 [4.59, 4.75] & 243\\
    \hline
  \end{tabular}
\end{adjustwidth}
\end{table}

As the exact size and shape of an Ego Network can vary somewhat between individuals, it can sometimes make it difficult to compare them with one another or across a social network as a whole.
To address this, it is common practice to standardise ENM analyses by focusing on Egos who have a number of circles exactly equal to 5 in their Ego Networks~\cite{Arnaboldi_2013, Dunbar_2015}.
The rightmost column of Table~\ref{tab:optimum_circle_and_mean_network_size} displays the number of such Egos for each dataset.
As the focus of this paper is most often on the active networks of all users, it shall be specifically stated whenever an analysis includes only Egos with 5 circles, as is the case or the circle-by-circle analysis that follows.

Next, the mean sizes of each individual circle, displayed in Table~\ref{tab:circle_sizes}, are observed. Recalling from Section~\ref{sec:introduction} that the outer circles of ENMs tend to be below standard in online contexts, one can see that the observed values are close to the expectations: i.e. 1-2, 5, 15, 45-50, 150. What's more, the scaling factor of 3 between each circle is evident, despite the smaller outer circles.
As with the overall network sizes, various external factors cause the outer circles to be marginally underpopulated~\cite{Arnaboldi_2013}.
Therefore, these values are in line with those of previous works on the ENM; the datasets appear to be appropriate for the computation of Ego Networks.
What's more, because the sizes of the circles and the active networks, as well as the mean number of circles, of all the datasets closely match those of the standard ENM, there is no discernible impact of culture or topic on these values.

\begin{table}
  \caption{Mean circle sizes of Egos with exactly 5 circles.}
  \label{tab:circle_sizes}
  \begin{tabular}{|l|l|c|c|c|c|c|}
    \hline
    \textbf{Dataset Type} & \textbf{Region} & \textbf{C\textsubscript{1}} & \textbf{C\textsubscript{2}} & \textbf{C\textsubscript{3}} & \textbf{C\textsubscript{4}} & \textbf{C\textsubscript{5}}\\
    \thickhline
    \multirow{1}{*}{Baseline} & -- & 1.78 & 6.16 & 16.86 & 44.19 & 125.91\\
    \hline
    \multirow{4}{*}{Geographical} & Mediterranean & 1.70 & 5.60 & 14.67 & 38.83 & 120.41\\
    & South America & 1.80 & 5.76 & 15.71 & 39.92 & 118.29\\
    & Northern Europe & 1.80 & 5.88 & 17.12 & 45.34 & 131.12\\
    & West Africa & 1.65 & 5.60 & 15.64 & 39.71 & 118.81\\ 
    \hline
    \multirow{3}{*}{Reality TV} & Italy & 1.63 & 5.15 & 13.85 & 35.60 & 103.65\\
    & Brazil & 1.58 & 4.65 & 12.08 & 31.29 & 96.63\\
    & Netherlands & 1.61 & 5.29 & 14.29 & 37.16 & 98.63\\
    \hline
    \multirow{3}{*}{Journalists} & Italy & 1.12 & 3.57 & 10.59 & 33.14 & 120.10\\
    & Brazil & 1.78 & 5.90 & 15.66 & 41.62 & 116.48\\
    & Netherlands & 1.66 & 5.51 & 15.54 & 43.08 & 122.69\\
    \hline
    \multirow{3}{*}{Generic Users} & Italy & 1.55 & 4.91 & 12.77 & 32.43 & 96.29\\
    & Brazil & 1.63 & 5.00 & 12.98 & 33.31 & 98.29\\
    & Netherlands & 1.62 & 5.09 & 13.25 & 33.76 & 97.11\\
    \hline
    \multirow{4}{*}{Weather} & Italy & 1.63 & 4.97 & 12.81 & 32.28 & 94.38\\
    & Brazil & 1.46 & 4.49 & 11.20 & 27.73 & 80.27\\
    & Netherlands & 1.34 & 4.29 & 10.41 & 25.80 & 75.96\\
    & Nigeria & 1.35 & 4.18 & 10.38 & 24.00 & 71.25\\
    \hline
    \multirow{4}{*}{Football} & Italy & 1.72 & 5.52 & 14.12 & 35.32 & 104.91\\
    & Brazil & 1.63 & 5.24 & 13.43 & 34.67 & 106.67\\
    & Netherlands & 1.56 & 4.84 & 12.27 & 29.51 & 86.15\\
    & Nigeria & 1.54 & 5.22 & 12.49 & 32.07 & 101.24\\
    \hline
    \multirow{4}{*}{Politics} & Italy & 1.70 & 5.38 & 13.40 & 33.38 & 101.22\\
    & Brazil & 1.70 & 5.07 & 13.31 & 33.11 & 97.55\\
    & Netherlands & 1.81 & 5.77 & 14.72 & 36.85 & 108.59\\
    & Nigeria & 1.52 & 4.56 & 11.47 & 27.70 & 83.22\\
    \hline
  \end{tabular}
  \vspace{-11pt}
\end{table}

Therefore, all of the collected networks display Egos with network structures that are aligned with the general findings of Ego Network research (see Section~\ref{sec:introduction}).

\subsection{Network Negativities by Topic and Region}
\label{sec:negs_by_topic_and_region}

Next, signs were computed for every relationship of every Ego in each dataset following the methodology detailed in Subsection~\ref{sec:signing_relationships}. This allows the mean negativity for each dataset to be easily calculated and compared.
To aid in this comparison, the datasets' overall mean negativities (including all Egos regardless of their number of circles) have been organised into two tables: Table~\ref{tab:user_negativities_old} for the pre-existing datasets and Table~\ref{tab:user_negativities_new} for the novel ones.
(Also, recall from Section~\ref{sec:datasets_old}, that the values of the datasets in the Geographical column of Table~\ref{tab:user_negativities_old} are not just the countries listed in the Region column but also those of culturally-similar, neighbouring countries).
The datasets in both tables are arranged into rows by region and into columns by type of user (Table~\ref{tab:user_negativities_old}) or topic (Table~\ref{tab:user_negativities_new}). Both rows and columns are then sorted by negativity.
When organised in this way, each negativity is larger than the one below it and to its left, with the only exceptions being Brazilian Journalists, Dutch Reality TV and Nigerian Weather. 
This strongly illustrates that both the geographical culture and the communities/topics that individuals are engaged with have a pronounced impact on the percentage of negative relationships that they maintain\footnote{It is worth noting that Italy and Brazil swap positions between the two tables. Suggesting that the order of geographical cultures based on negativity can vary slightly between user types and/or topics.}.
What's more, a clear pattern naturally emerges in the order of the topics in Table~\ref{tab:user_negativities_new}: from least polarising (Generic) to most polarising (politics).
This illustrates that the more polarising a topic is, the more negative relationships users engaging with it have, and this holds across all cultures.
Additionally, the negativities of the country-specific Generic Users are very close to those of the Geographical datasets, which also include users from neighbouring countries. This would suggest users from the countries that have been grouped together in the latter are fairly similar in terms of negative relationship percentage.

\begin{table}
  \caption{Mean user negativities for datasets taken from previous papers, arranged by region and user type and ordered by negativity. Ranges between the Brazilian, Italian and Dutch datasets, ranges between all the datasets and ranges between all the topics are also displayed.}
  \label{tab:user_negativities_old}
  \begin{tabular}{|l|c|c|c|c|}
    \hline
    \textbf{Region} & \textbf{Geographical} & \textbf{Journalists} & \textbf{Reality TV} & \textbf{Range}\\
    \thickhline
    Brazil & 65.67\textsuperscript{1} & 64.93 & 69.47 & 4.89\\
    Italy & 60.08\textsuperscript{1} & 63.87 & 64.97 & 4.54\\
    Netherlands & 54.66\textsuperscript{1} & 57.65 & 68.36 & 13.71\\
    Nigeria & 50.29\textsuperscript{1} & -- & -- & --\\
    Baseline & 49.90 & -- & -- & --\\
    \hline
    Range\textsubscript{BIN} & 11.01 & 7.27 & 4.50 & --\\
    Range\textsubscript{all} & 15.78 & 7.27 & 4.50 & --\\
    \hline
  \end{tabular}
  \\
  \footnotesize{\textsuperscript{1}Note: The Geographical datasets are not exclusively the counties listed in the Region column, but also include a few neighbouring countries (see Subsection~\ref{sec:datasets_novel})}
\end{table}

\begin{table}
  \caption{Mean user negativities for datasets collected for this paper, arranged by region and topic and ordered by negativity. Ranges between the Italian, Brazilian and Dutch datasets, ranges between all the datasets and ranges between all the topics are also displayed.}
  \label{tab:user_negativities_new}
  \begin{tabular}{|l|c|c|c|c|c|}
    \hline
    \textbf{Region} & \textbf{Generic} & \textbf{Weather} & \textbf{Football} & \textbf{Politics} & \textbf{Range}\\
    \thickhline
    Italy & 66.75 & 74.24 & 77.96 & 84.34 & 10.10\\
    Brazil & 64.58 & 69.88 & 71.41 & 81.24 & 11.36\\
    Netherlands & 54.55 & 56.67 & 64.09 & 78.25 & 21.57\\
    Nigeria & -- & 60.73 & 58.45 & 68.30 & 9.86\\
    \hline
    Range\textsubscript{IBN} & 12.19 & 17.57 & 13.87 & 6.09 & --\\
    Range\textsubscript{all} & 12.19 & 17.57 & 19.51 & 16.03 & --\\
    \hline
  \end{tabular}
\end{table}

While the results of the previous paragraph are unsurprising given the results of previous works (namely~\cite{Tacchi_2023}), what is surprising are the ranges of the novel datasets. 
The range column in both tables shows the range of negativity across different communities for a same country. 
Conversely, the range rows show the ranges across different countries for the same type of community. 
Note that, to better compare the values of the two tables, an additional range row was computed for only countries that have complete datasets in the new datasets, (i.e. all but Nigeria).
As mentioned in Section~\ref{sec:introduction}, it has been observed that cultural differences in negativity tend to be overpowered by the influence of the communities with which we are involved and that this effect is stronger the more negative or polarising a social group or topic is.
Indeed, this effect appears visible in the range rows of Table~\ref{tab:user_negativities_old}: the more negative types of users display lower ranges (i.e. less difference between the different cultures).
However, this effect is not visible for the novel datasets, where the ranges seem to oscillate without any correlation with this expectation. Instead, the ranges appear to not be strictly determined by the topics.
However, focusing on the ranges for the 3 countries that have available datasets in each novel category (i.e. excluding Nigeria), there is a topic with a lower range than the Generic dataset: Politics. 
This, of course, is the most polarising of all the datasets collected, which may indicate that cultural effects may only be suppressed by topics that are exceedingly polarising. 
In the datasets of Table~\ref{tab:user_negativities_old}, the same holds for Reality TV, which can also be considered as a topic eliciting quite confrontational discussions.
Thus, it appears that the impact of culture is not always being ``overwhelmed'' by that of topics or subcommunities with which an individual is engaged as previously thought~\cite{Tacchi_2023}. 
Rather the strength of the influence a given topic or community has on an individual depends somewhat on their culture and that, although topics can overwhelm cultural differences in extreme cases (e.g. Politics), this is not as common as previously thought.
Intuitively, this conclusion seems logical as values and priorities can change dramatically between cultures, whereas politics can lead to strong and contradicting opinions in almost any culture. 
Indeed, the ranges could be taken as a measurement of how much the value of a given topic varies from culture to culture, rather than an inverse measure of how polarising it is. 
Therefore, future research may want to take steps to consider which topics could be considered polarising for each culture separately.

In contrast to those of the topics, the ranges of the countries (range columns) are more consistent: with all countries except the Netherlands displaying ranges that are relatively close to one another in both tables. 
This suggests that, while the impact a topic has on negativity across a set of cultures can vary drastically, the cultural influence on negativity across a set of topics may be, at least somewhat, predictable.

\subsection{Signed Ego Network Analysis}
\label{sec:signed_ego_network_analysis}
Next, the distribution of the relationship signs across each circle is observed.
As with the unsigned analysis of the circles, here the focus is on Egos with exactly 5 circles. 
The mean numbers and percentages of negative relationships for each of the 5 circles are displayed in Table~\ref{tab:circle_negativities}.
The most negative circle of each dataset is emphasised in bold. Thus, one can see that Circle 1 is the most negative circle for 7 of the datasets, Circle 2 for 13 of them and Circle 3 for the remaining 6.
This is in line with previous findings on SENMs, as the inner circles are often found to contain the highest density of negative relationships~\cite{Tacchi_2022}.

Similarly in line with previous findings is that negativity percentages tend to be higher across all circles for datasets which are centred around a more specific topic. For example, the circle negativities of the Italian Generic Users are 67.78, 70.40, 69.50, 66.58 and 62.18 and each of these negativities is less than that of the corresponding circle for the Italian Weather dataset, 76.99, 76.66, 79.68, 78.87 and 74.87, which in turn are lower than the corresponding Italian Football negativities, 84.80, 87.84, 87.75, 86.40, 79.50, and so on.
Therefore, there is a clearly visible increase in negativity for more specific topics and this increase is observable for all cultures and at all levels of the SENM.

Observing the variation in negativity across the first 4 circles\footnote{Circle 5 is not included because it has previously been found that Alters in this circle usually have an average number of interactions of between 3 and 4, which is not considered reliable given the 1:5 ratio of the Golden Interaction Ratio (described in Section
~\ref{sec:introduction}) used to sign the relationships~\cite{Tacchi_2024}.} and the corresponding range reveals further insights.
It has previously been observed that the percentages of negative relationships change less from circle to circle for users related to more generic sets of topics, as opposed to users related to more specific topics~\cite{Tacchi_2023}.
Otherwise said, generic users tend to have negative relationships that are evenly distributed across their circles, whereas more specialised users often have more higher numbers of negative relationships in their inner circles.
This can be seen for the previous datasets, where the Baseline and Geographical datasets have, with one exception (Northern Europe), ranges between 3 and 5. 
Conversely, more specific datasets tended to have larger ranges, all of which are above 5 and half of which are over 10.
It has been hypothesised that the increased ranges may be due to increased levels of user engagement and/or more polarising topics, resulting in a greater diversity of relationship negativities~\cite{Tacchi_2023}.
For the novel datasets, the previous findings are replicated for the generic dataset, Generic Users and Weather, which all have ranges between 2 and 6. 
However, what is more surprising is that the ranges of the specific datasets, Football and Politics, also fall within this interval (the only exception to this is Nigerian Football). 
Furthermore, Politics, which would be expected to have the most variation between circles, instead, as was the case for the ranges between cultures in Subsection~\ref{sec:negs_by_topic_and_region}, show quite small ranges. 
This further supports the suggestion that very polarising topics can overpower other differences in negativities, causing them to converge, between cultures as previously seen, and also between different individuals within the same culture, as seen here.

These latter observations would suggest that contrary to previous conclusions, user type does not have an observable impact on the variations in negativity across circles.
However, as mentioned when discussing the ranges of the overall negativities of each dataset (Subsection~\ref{sec:negs_by_topic_and_region}), how polarising a given topic is may vary significantly between different cultures. This may be an interesting avenue for future research.

\begin{sidewaystable}
  \caption{Mean number of negative relationships for Egos with 5 circles for each circle, with percentages in parentheses and the most negative circle of each dataset in bold. The percentage range between C\textsubscript{1} and C\textsubscript{4} is also displayed.}
  \label{tab:circle_negativities}
  \begin{tabular}{|l|l|c|c|c|c|c|c|}
    \hline
    \textbf{Dataset Type} & \textbf{Region} & \textbf{C\textsubscript{1}} & \textbf{C\textsubscript{2}} & \textbf{C\textsubscript{3}} & \textbf{C\textsubscript{4}} & \textbf{C\textsubscript{5}} & \textbf{Range\textsubscript{C\textsubscript{1}C\textsubscript{4}}}\\
    \thickhline
    Baseline & -- & 1.00 (56.25\%) & \textbf{3.63 (58.84\%)} & 9.72 (57.64\%) & 24.28 (54.95\%) & 63.71 (50.60\%) & 3.90\\
    \hline
    \multirow{4}{*}{Geographical} & Mediterranean & \textbf{1.25 (73.58\%)} & 4.06 (72.54\%) & 10.38 (70.77\%) & 27.07 (69.70\%) & 76.85 (63.82\%) & 3.88\\
    & South America & 1.37 (76.42\%) & \textbf{4.42 (76.76\%)} & 12.00 (76.38\%) & 28.71 (71.93\%) & 75.03 (63.43\%) & 4.83\\
    & Northern Europe & \textbf{1.26 (69.86\%)} & 3.94 (67.05\%) & 11.04 (64.48\%) & 27.45 (60.54\%) & 70.67 (53.89\%) & 9.32\\
    & West Africa & 0.92 (55.40\%) & \textbf{3.18 (56.81\%)} & 8.75 (55.94\%) & 21.25 (53.51\%) & 60.80 (51.17\%) & 3.30\\
    \hline
    \multirow{3}{*}{Reality TV} & Italy & 1.08 (66.15\%) & 3.85 (74.76\%) & \textbf{10.58 (76.35\%)} & 26.38 (74.09\%) & 71.38 (68.86\%) & 10.20\\
    & Brazil & \textbf{1.31 (82.89\%)} & 3.83 (82.51\%) & 9.92 (82.07\%) & 24.31 (77.70\%) & 67.73 (70.09\%) & 5.20\\
    & Netherlands & 1.15 (71.00\%) & 3.97 (75.00\%) & \textbf{10.97 (76.75\%)} & 27.73 (74.61\%) & 67.42 (68.36\%) & 5.75\\
    \hline
    \multirow{3}{*}{Journalists} & Italy & \textbf{1.00 (89.47\%)} & 3.12 (87.36\%) & 8.67 (81.85\%) & 25.80 (77.87\%) & 84.80 (70.61\%) & 11.60\\
    & Brazil & 1.14 (64.04\%) & 4.42 (74.92\%) & \textbf{11.94 (76.25\%)} & 30.20 (72.56\%) & 77.02 (66.12\%) & 12.20\\
    & Netherlands & \textbf{1.19 (71.60\%)} & 3.90 (70.74\%) & 10.78 (69.34\%) & 28.13 (65.30\%) & 71.50 (58.27\%) & 6.31\\
    \hline
    \multirow{3}{*}{Generic Users} & Italy & 1.05 (67.78\%) & \textbf{3.45 (70.40\%)} & 8.87 (69.50\%) & 21.59 (66.58\%) & 59.87 (62.18\%) & 3.82\\
    & Brazil & 1.19 (72.65\%) & \textbf{3.68 (73.72\%)} & 9.48 (73.04\%) & 23.43 (70.72\%) & 63.15 (64.33\%) & 3.00\\
    & Netherlands & 0.93 (56.97\%) & 3.01 (59.12\%) & \textbf{7.87 (59.39\%)} & 19.44 (57.59\%) & 51.82 (53.36\%) & 2.42\\
    \hline
    \multirow{4}{*}{Weather} & Italy & 1.25 (76.88\%) & 3.81 (76.66\%) & \textbf{10.21 (79.68\%)} & 25.46 (78.87\%) & 70.66 (74.87\%) & 3.02\\
    & Brazil & 1.18 (81.16\%) & \textbf{3.65 (81.32\%)} & 8.88 (79.26\%) & 21.42 (77.27\%) & 56.26 (70.09\%) & 4.05\\
    & Netherlands & 0.84 (62.67\%) & \textbf{2.79 (65.00\%)} & 6.70 (64.32\%) & 16.20 (62.77\%) & 45.14 (59.43\%) & 2.33\\
    & Nigeria & \textbf{1.01 (75.00\%)} & 3.10 (74.30\%) & 7.54 (72.66\%) & 16.74 (69.73\%) & 45.57 (63.96\%) & 5.27\\
    \hline
    \multirow{4}{*}{Football} & Italy & 1.46 (84.80\%) & \textbf{4.85 (87.84\%)} & 12.39 (87.75\%) & 30.51 (86.40\%) & 83.41 (79.50\%) & 3.04\\
    & Brazil & 1.34 (82.28\%) & \textbf{4.34 (82.81\%)} & 10.92 (81.31\%) & 27.13 (78.25\%) & 74.55 (69.89\%) & 4.56\\
    & Netherlands & 1.05 (67.13\%) & \textbf{3.43 (70.90\%)} & 8.69 (70.84\%) & 20.35 (68.97\%) & 54.65 (63.44\%) & 3.77\\
    & Nigeria & 1.20 (77.78\%) & \textbf{4.12 (78.97\%)} & 8.95 (71.68\%) & 21.02 (65.55\%) & 59.51 (58.78\%) & 13.42\\
    \hline
    \multirow{4}{*}{Politics} & Italy & 1.56 (91.82\%) & \textbf{4.95 (92.01\%)} & 12.25 (91.43\%) & 29.94 (89.70\%) & 84.98 (83.95\%) & 2.31\\
    & Brazil & 1.54 (90.79\%) & 4.71 (92.79\%) & \textbf{12.36 (92.88\%)} & 29.99 (90.56\%) & 80.57 (82.60\%) & 2.32\\
    & Netherlands & 1.54 (85.14\%) & \textbf{5.03 (87.13\%)} & 12.78 (86.78\%) & 31.09 (84.38\%) & 84.89 (78.17\%) & 2.75\\
    & Nigeria & \textbf{1.16 (76.69\%)} & 3.47 (76.01\%) & 8.70 (75.84\%) & 20.57 (74.26\%) & 57.11 (68.62\%) & 2.44\\
    \hline
  \end{tabular}
\end{sidewaystable}

\subsection{Topic Coherence and Diversity}
\label{sec:coherence_and_diversity}

In Subsection~\ref{sec:topic_analysis}, BERTopic is used to identify the most popular topics in the 3 large, Generic User datasets.
However, before using BERTopic, it is important to ensure that its parameters are appropriately tuned for the data being used.
For this, 2 metrics can be utilised to gauge how congruent BERTopic's generated topics are (Coherence) as well as how much variation there is between them (Diversity).

The Coherence is calculated as the mean Normalised Pointwise Mutual Information (NPMI) of two words drawn randomly from the same document, defined between 0 and 1~\cite{Dieng_2020}.
Coherence scores can vary greatly based on the nature of the data being used, and what is a good score in one context may be a bad score in another~\cite{Rosner_2014}.
Previous research has reported coherence scores of between 0.438 and 0.806 for a collection of articles from Wikipedia, the New York Times and other similar structured data sources~\cite{Roder_2015}.
With the comparatively messy nature of X data, one might reasonably expect its coherence scores to be lower and, indeed, this does appear to be the case.
Coherence scores using datasets collected using X hashtags found coherence scores between 0.127 and 0.205~\cite{Fang_2015}.
Given the snowball sampling collection method of the datasets used here (as opposed to using hashtags), the expected values could even be slightly lower than this.
The Coherence scores can be seen in Table~\ref{tab:coherence_scores}.

The Diversity is the percentage of unique words in the top 25 words of all topics~\cite{Dieng_2020}.
It is also defined between 0 and 1, with 0 indicating many repeating or superfluous topics and 1 signifying many unique and varied topics.
The Diversity scores of the novel datasets can be seen in Table~\ref{tab:diversity_scores}.

\begin{table}
  \caption{The Coherence scores of the 15 novel datasets.}
  \label{tab:coherence_scores}
  \begin{tabular}{|l|c|c|c|c|}
    \hline
    \textbf{Region} & \textbf{Generic} & \textbf{Weather} & \textbf{Football} & \textbf{Politics}\\
    \thickhline
    Italy & 0.134 & 0.134 & 0.140 & 0.225\\
    Brazil & 0.116 & 0.117 & 0.179 & 0.168\\
    Netherlands & 0.200 & 0.139 & 0.130 & 0.206\\
    Nigeria & -- & 0.180 & 0.241 & 0.290\\
    \hline
  \end{tabular}
\end{table}

\begin{table}
  \caption{The Diversity scores of the 15 novel datasets.}
  \label{tab:diversity_scores}
  \begin{tabular}{|l|c|c|c|c|}
    \hline
    \textbf{Region} & \textbf{Generic} & \textbf{Weather} & \textbf{Football} & \textbf{Politics}\\
    \thickhline
    Italy & 0.930 & 0.934 & 0.913 & 0.904\\
    Brazil & 0.907 & 0.936 & 0.918 & 0.926\\
    Netherlands & 0.905 & 0.936 & 0.920 & 0.905\\
    Nigeria & -- & 0.900 & 0.915 & 0.860\\
    \hline
  \end{tabular}
\end{table}

Observing Table~\ref{tab:coherence_scores}, one can see that the Coherence scores are all within the range 0.116 and 0.241, which is similar to the 0.127 and 0.205 range previously reported using X datasets~\cite{Fang_2015}.
Thus, these values seem to indicate a logically good connections between the resulting topics of each dataset.
While the scores are similar for 3 of the 4 countries, with Nigeria being somewhat higher.
Similarly, there is a clear increase from Generic, to Weather, to Football, to Politics, for all 4 countries (with the exception of the Netherlands).
Due to the snowball sampling methodology by which the data were collected, it was not possible to guarantee that a majority of users were engaging meaningfully in the target topic of their dataset.
However, the higher Coherence scores for Football and Politics would suggest users' conversation points become more focused when the topic of discussion is more polarising; supporting the assumption that users' in these datasets are engaging meaningfully in their target topic.
Although, given the small sample size and the differing values of the Dutch datasets, this should be taken with a grain of salt.

Similarly, Table~\ref{tab:diversity_scores} displays very high levels of diversity, with almost all of datasets displaying above 0.9.
This suggests an extremely high level of coverage of the topics generated by BERTopic.
In contrast to Coherence, here there are only minor variations between the different countries and dataset types, for example the Nigerian datasets' Diversity scores are slightly lower than the others.

Taking the Coherence and Diversity scores together, the BERTopic-generated topics appear to be both wide ranging and reasonably coherent given the source of the data.
Therefore, both the datasets and the parameters of BERTopic appear to be appropriate for the topic analysis in the following subsection.

\subsection{Topic Analysis}
\label{sec:topic_analysis}

Given the clear differences in rates of negative relationships, resulting from the influence of topics, a further analysis was conducted to see whether dividing a generic dataset into some of its constituent topics would yield the same findings.
As mentioned in Subsection~\ref{sec:method_topic_analysis}, BERTopic can be used to obtain a list of key terms that are being used for each of these topics, making it possible to check each individual tweet to see if it is related to any of the top topics. 
Specifically, it was used to obtain the main topics being discussed in each of the 3 novel Generic Users datasets (those for Italy, Brazil and the Netherlands). 
This led to a list of IDs of both the tweets and the users involved, which in turn allowed the topics to be matched to 2 negativity metrics. 
The first of these is the percentage of negative relationships within the Ego Network of users that tweeted in relation to each topic, taken as the mean of all the users. 
The resulting value provides a gauge the negative impact a topic has on the relationships of those who are engaged with it. 
The second metric is simply the percentage of related tweets which are negative, which reveals how negative each topic is in isolation, i.e. irrespective of the surrounding network. 
For both of these metrics, the final negativity is the average of the top 3 terms for each topic, weighted by number of occurrences.
The metrics are displayed in Table~\ref{tab:top_topics_user_negativities} (mean percentage of negative relationships) and Table~\ref{tab:top_topics_tweet_negativities} (percentage of negative tweets); to save space, only the top term is shown for each topic.
In addition, some specific categories of topics have been chosen to be focused on in more detail.
These categories represent a mix between the specific topics of the other datasets collected for this paper (Politics, Football and Generic) as well as two additions: COVID, which provides a unique opportunity to analyse a single event which has had an impact on every single person across the globe, and Religion, which was noted as a specific topic of interest for future research in previous work~\cite{Tacchi_2023}. 
Both the aforementioned tables have been colour-coded to make these categories more visible.

\begin{table}
\begin{adjustwidth}{-0.45in}{0in}
  \caption{Top 20 topics for the 3 Generic Users datasets and the mean percentage of negative relationships of users who are engaged with each topic, ordered by negativity. The topics are colour-coded: red for Politics, green for COVID, yellow for Religion, purple for Football and blue for Generic.}
  \label{tab:top_topics_user_negativities}
  \begin{tabular}{|l+r|c+r|c+r|c|}
    \hline
     & \multicolumn{2}{c+}{\textbf{Italy (66.75\%)}} & \multicolumn{2}{c+}{\textbf{Brazil (64.58\%)}} & \multicolumn{2}{c|}{\textbf{Netherlands (54.55\%)}}\\
     \hline
    \textbf{Index} & \textbf{Topic} & \textbf{Negativity} & \textbf{Topic} & \textbf{Negativity} & \textbf{Topic} & \textbf{Negativity}\\
    \thickhline
    1 & \cellcolor{political!60}salvini & 89.17 & \cellcolor{political!60}lava & 81.03 & \cellcolor{political!60}biden & 64.03\\
    2 & \cellcolor{political!60}renzi & 86.88 & \cellcolor{political!60}putin & 80.50 & \cellcolor{political!60}haag & 61.42\\
    3 & \cellcolor{generic!80}amazon & 79.07 & \cellcolor{political!60}gasolina & 76.93 & \cellcolor{generic!80}lachen & 60.05\\
    4 & \cellcolor{generic!80}peggio & 77.89 & \cellcolor{generic!80}menino & 76.49 & \cellcolor{generic!80}nope & 59.96\\
    5 & \cellcolor{generic!80}odio & 76.97 & \cellcolor{generic!80}verdades & 75.93 & \cellcolor{generic!80}hond & 58.77\\
    6 & \cellcolor{political!60}democrazia & 76.93 & \cellcolor{religious}jesus & 75.53 & \cellcolor{generic!80}gemist & 57.07\\
    7 & \cellcolor{generic!80}nomi & 75.03 & \cellcolor{generic!80}cabelo & 75.45 & \cellcolor{generic!80}trein & 56.42\\
    8 & \cellcolor{generic!80}pizza & 74.75 & \cellcolor{generic!80}perdi & 74.88 & \cellcolor{generic!80}vakantie & 55.90\\
    9 & \cellcolor{covid!70}virus & 74.26 & \cellcolor{generic!80}fã & 73.44 & \cellcolor{generic!80}slapen & 55.74\\
    10 & \cellcolor{religious}papa & 74.17 &  \cellcolor{generic!80}festa & 73.00 & \cellcolor{covid!70}coronavirus & 54.74\\
    11 & \cellcolor{generic!80}concordo & 73.71 & \cellcolor{generic!80}meme & 72.77 & \cellcolor{generic!80}filmpje & 53.18\\
    12 & \cellcolor{football!60}calcio & 72.99 &  \cellcolor{covid!70}máscara & 72.01 & \cellcolor{generic!80}gold & 51.53\\
    13 & \cellcolor{generic!80}thread & 69.35 &  \cellcolor{generic!80}netflix & 71.13 & \cellcolor{generic!80}aflevering & 51.15\\
    14 & \cellcolor{generic!80}dibattito & 68.54 &  \cellcolor{generic!80}barato & 70.99 & \cellcolor{generic!80}facebook & 50.93\\
    15 & \cellcolor{generic!80}caffè & 66.71 &  \cellcolor{generic!80}gato & 68.91 & \cellcolor{generic!80}seizoen & 50.38\\
    16 & \cellcolor{generic!80}natale & 66.47 &  \cellcolor{generic!80}pizza & 67.54 & \cellcolor{generic!80}koffie & 49.78\\
    17 & \cellcolor{generic!80}sogno & 66.28 &  \cellcolor{generic!80}facebook & 67.52 & \cellcolor{generic!80}verjaardag & 48.69\\
    18 & \cellcolor{covid!70}coronavirus & 66.25 &  \cellcolor{generic!80}artista & 66.11 & \cellcolor{generic!80}interviews & 48.40\\
    19 & \cellcolor{generic!80}facebook & 63.84 &  \cellcolor{generic!80}natal & 64.99 & \cellcolor{generic!80}fotos & 46.71\\
    20 & \cellcolor{generic!80}serata & 62.33 &  \cellcolor{generic!80}dm & 63.96 & \cellcolor{generic!80}anniversary & 40.22\\
    \hline
  \end{tabular}
\end{adjustwidth}
\end{table}

\begin{table}
\begin{adjustwidth}{-0.45in}{0in}
  \caption{Top 20 topics for the 3 Generic Users datasets and the mean negativity of all corresponding tweets, ordered by negativity. The topics are colour-coded: red for Politics, green for COVID, yellow for Religion, purple for Football and blue for Generic.}
  \label{tab:top_topics_tweet_negativities}  
  \begin{tabular}{|l+r|c+r|c+r|c|}
    \hline
     & \multicolumn{2}{c+}{\textbf{Italy (66.75\%)}} & \multicolumn{2}{c+}{\textbf{Brazil (64.58\%)}} & \multicolumn{2}{c|}{\textbf{Netherlands (54.55\%)}}\\
     \hline
    \textbf{Index} & \textbf{Topic} & \textbf{Negativity} & \textbf{Topic} & \textbf{Negativity} & \textbf{Topic} & \textbf{Negativity}\\
    \thickhline
    1 & \cellcolor{generic!80}peggio & 95.33 & \cellcolor{political!60}putin & 85.59 & \cellcolor{generic!80}gemist & 55.59 \\
    2 & \cellcolor{generic!80}odio & 90.09 & \cellcolor{generic!80}perdi & 82.11 & \cellcolor{covid!70}coronavirus & 45.90 \\
    3 & \cellcolor{covid!70}virus & 83.03 & \cellcolor{political!60}lava & 68.50 & \cellcolor{political!60}haag & 45.15 \\
    4 & \cellcolor{political!60}salvini & 81.18 & \cellcolor{religious}jesus & 59.87 & \cellcolor{generic!80}hond & 41.87 \\
    5 & \cellcolor{political!60}renzi & 80.94 & \cellcolor{covid!70}máscara & 59.01 & \cellcolor{generic!80}trein & 37.23 \\
    6 & \cellcolor{political!60}democrazia & 60.09 & \cellcolor{generic!80}verdades & 49.71 & \cellcolor{political!60}biden & 36.02 \\
    7 & \cellcolor{covid!70}coronavirus & 59.40 & \cellcolor{political!60}gasolina & 48.55 & \cellcolor{generic!80}lachen & 34.38 \\
    8 & \cellcolor{religious}papa & 54.53 & \cellcolor{generic!80}menino & 47.64 & \cellcolor{generic!80}slapen & 34.31 \\
    9 & \cellcolor{football!60}calcio & 53.02 & \cellcolor{generic!80}meme & 41.45 & \cellcolor{generic!80}nope & 29.83 \\
    10 & \cellcolor{generic!80}nomi & 51.67 & \cellcolor{generic!80}gato & 40.71 & \cellcolor{generic!80}vakantie & 25.66 \\
    11 & \cellcolor{generic!80}pizza & 47.01 & \cellcolor{generic!80}cabelo & 37.08 & \cellcolor{generic!80}facebook & 25.46 \\
    12 & \cellcolor{generic!80}amazon & 43.00 & \cellcolor{generic!80}festa & 36.05 & \cellcolor{generic!80}filmpje & 23.18 \\
    13 & \cellcolor{generic!80}concordo & 38.53 & \cellcolor{generic!80}fã & 35.97 & \cellcolor{generic!80}gold & 19.07 \\
    14 & \cellcolor{generic!80}dibattito & 33.88 & \cellcolor{generic!80}facebook & 35.67 & \cellcolor{generic!80}seizoen & 16.92 \\
    15 & \cellcolor{generic!80}caffè & 32.23 & \cellcolor{generic!80}barato & 35.27 & \cellcolor{generic!80}fotos & 16.88 \\
    16 & \cellcolor{generic!80}facebook & 30.36 & \cellcolor{generic!80}netflix & 34.91 & \cellcolor{generic!80}interviews & 16.75 \\
    17 & \cellcolor{generic!80}natale & 28.42 & \cellcolor{generic!80}pizza & 32.03 & \cellcolor{generic!80}koffie & 16.13 \\
    18 & \cellcolor{generic!80}thread & 26.81 & \cellcolor{generic!80}artista & 30.95 & \cellcolor{generic!80}aflevering & 11.59 \\
    19 & \cellcolor{generic!80}sogno & 24.39 & \cellcolor{generic!80}natal & 22.58 & \cellcolor{generic!80}anniversary & 10.16 \\
    20 & \cellcolor{generic!80}serata & 11.38 & \cellcolor{generic!80}dm & 16.74 & \cellcolor{generic!80}verjaardag & 8.72 \\
    \hline
  \end{tabular}
\end{adjustwidth}
\end{table}

Comparing the two negativity metrics provides a deeper understanding of how each topic affects negative relationships. 
For example, the words ``peggio'' and ``gemist'' are keywords in the topics most likely to be in a negative tweet, for the Italian and Dutch Generic Users datasets respectively (Table~\ref{tab:top_topics_tweet_negativities}). 
This is unsurprising given that they mean ``worse'' or ``the worst'' in Italian and ``excrement'' in Dutch. 
However, they drop down to the 4th and 6th most negative topics in terms of impact on a user's relationships (Table~\ref{tab:top_topics_user_negativities}). 
Revealing that, although they are frequently used in negative contexts, their influence in terms of relationships is lower than expected. By comparison, ``salvini'' and ``biden'', both well-known politicians, have the greatest negative impact on users' relationships out of any topics of the Italian and Dutch datasets while being at the 4th and 6th positions for tweet negativity. This shows that, although they are less frequently used in negative contexts than ``peggio'' and ``gemist'', they are much stronger indicators of negativity in a user's surrounding network.

Subsequently, the negativities of the top topics are viewed when grouped into the aforementioned categories (Politics, Religion, Football, COVID and Generic). 
As was just touched upon, Politics is by far the most negative category of topic in all three datasets. They are also the most specific topics, with all of the topics referencing specific people (Matteo Salvini, Matteo Renzi, Vladimir Putin and Joe Biden), places (The Hague) or problems (Operação Lava Jato and rising fuel prices/Petroleo Brasileiro SA). The only non-specific political topic is ``democrazia'' (``democracy'' in Italian), which is also the least negative (in terms of placement). Thus, the higher negativity of the political category of topics may be due to their specificity, which, as previously mentioned, can lead to a greater number of conflicting opinions.

Next, the two smallest categories: Religion and Football. Topics relating to these categories only appeared in the Italian and Brazilian datasets for the former and in the Italian for the latter. The religious topics were focused on specific individuals (the pope and Jesus) whereas Football is a generic topic relating to the sport as a whole (``calcio'' being the Italian for football). Despite the religious terms being just as specific as the political ones, they are visibly less negative overall, suggesting that specificity alone is not enough to explain the differing negativities. Football appears just below the corresponding religious topic in its dataset in both of the tables. 

The COVID category shows the biggest difference between the two tables. In Table~\ref{tab:top_topics_user_negativities} these topics appear towards the lower half of the table and, the least negative non-Generic term in each of the three datasets belongs to the COVID category. 
However, in Table~\ref{tab:top_topics_tweet_negativities}, it is almost the exact opposite, with the most negative non-Generic term belonging to this category for both the Italian and Dutch datasets. 
This difference between the tables may suggest that, while individual Tweets related to COVID do tend to be very negative, tweeting negatively about COVID does not have an overly strong negative effect on an individual's surrounding relationships. 
This highlights the difference between a negative topic and a polarising one. 
Despite COVID being a very negative category, there appear to be fewer differing opinions about it (i.e. there is a fairly strong consensus that COVID overall is bad) and, therefore, fewer disagreements compared to the other selected categories.

Counting the total number of topics in the final category, Generic, for each dataset, Italy has the fewest (13), followed by Brazil (15) and then the Netherlands (17). 
These numbers are reversely correlated with the order of the overall negativities of these datasets, which are 66.75\%, 64.58\% and 54.55\% respectively. 
Thus, it appears that the more specific topics a user is engaged with, the higher the percentage of negative relationships they have is expected to be.
While this finding supports previous work, which came to the same conclusion~\cite{Tacchi_2023}, it is important to note that given the small sample size, further work would need to be carried out before any conclusions can be made definitively.

Finally, the categories of topics were grouped together and their mean values were calculated in terms of both tweet negativity (Table~\ref{tab:top_topics_tweet_negativities}) and user negativity (Table~\ref{tab:top_topics_user_negativities}). These values are displayed in Table~\ref{tab:mean_category_negs}.

\begin{table}
  \caption{Mean user and tweet negativities, by category, of the top 20 topics in the 3 novel Generic Users datasets, ordered by negativity. The categories are colour-coded: red for Politics, green for COVID, yellow for Religion, purple for Football and blue for Generic.}
  \label{tab:mean_category_negs}  
  \begin{tabular}{|c|c+c|c+c|c|}
    \hline
     \multicolumn{2}{|c+}{\textbf{Italy}} & \multicolumn{2}{c+}{\textbf{Brazil}} & \multicolumn{2}{c|}{\textbf{Netherlands}}\\
     \hline
    \textbf{User} & \textbf{Tweet} & \textbf{User} & \textbf{Tweet} & \textbf{User} & \textbf{Tweet}\\
    \thickhline
    \cellcolor{political!60}84.32 & \cellcolor{political!60}74.07 & \cellcolor{political!60}79.48 & \cellcolor{political!60}67.55 & \cellcolor{political!60}62.73 & \cellcolor{covid!70}45.90\\
    \cellcolor{religious}74.17 & \cellcolor{covid!70}71.22 & \cellcolor{religious}75.53 & \cellcolor{religious}59.87 & \cellcolor{covid!70}54.74 & \cellcolor{political!60}40.59\\
    \cellcolor{football!60}72.99 & \cellcolor{religious}54.53 & \cellcolor{covid!70}72.01 & \cellcolor{covid!70}59.01 & \cellcolor{generic!80}52.64 & \cellcolor{generic!80}26.10\\
    \cellcolor{generic!80}70.84 & \cellcolor{football!60}53.02 & \cellcolor{generic!80}70.87 & \cellcolor{generic!80}38.24 & -- & --\\
    \cellcolor{covid!70}70.25 & \cellcolor{generic!80}39.91 & -- & -- & -- & --\\
    \hline
  \end{tabular}
\end{table}

Similar to the previous results, Politics is the most negative overall and is the most negative for all columns except the Dutch Tweet negativity, where it is beaten by COVID. 
Religion is usually the second most negative for the two datasets in which it appears, Italy and Brazil, and it is followed by Football in the former.
COVID is the least consistent, appearing in 4 of the 5 possible positions.
Finally, Generic is the least negative, appearing in the least negative position for all columns except the Italian user negativity, where it appears above COVID.

As these are the means of other values previously discussed in detail, it is unsurprising that they do not provide any additional results. 
Their main value is to confirm the overall negativity of very polarising topics (such as Politics), and the lower negativity of neutral topics (Generic) with respect to the more specific topics (Football, Religion, COVID).

\section{Conclusion}
\label{sec:conclusion}

Overall, this paper provides a thorough analysis of relationship negativities between cultures and topics in a systemised manner using 26 datasets, 15 of which were especially collected for this task. These datasets were specifically gathered to obtain a more precise understanding of previously observed phenomena. This paper also builds on previous works to provide a significantly improved understanding of how negativities can be affected at each level of the SENM as well as at the level of the overall network. The results provide further support for some previous hypotheses (e.g. more specific or polarising topics lead to greater negativities at all levels of the SENM) and redefine others (e.g. how easily cultural effects are overwhelmed by topics).

Specifically, the main take-home messages of the analysis presented in this paper are: (i) there appear to be some differences in negativity due to both culture and topic. 
However, cultural differences do not seem to get ``overwhelmed'' by topics as easily as previously thought, rather, the impact of culture on negativity is important in the majority of cases and is only overpowered by very polarising topics; (ii) networks centred around a specific topic display a dramatic increase in negativity across all cultures and this effect is stronger the more negative or polarising the topic; (iii) the stability of negativity across the different circles of the SENM does not seem to decrease for more negative or specific topics, as was found previously; (iv) the number of polarising topics that appear among those most talked about within a dataset is a good indicator of how negative (in terms of relationships) that dataset is, and this is true even when tweets relating to those topics aren't overly negative themselves.

As with any piece of research, there are some important limitations to bear in mind while considering the results. 
For instance, the SVM method used to filter non-human users (mentioned in Section~\ref{sec:non-human_users}) has reasonably high reported accuracy for detecting the more simplistic types of bots that make up the majority of bots on Twitter~\cite{Lopez-Joya_2023}.
However, it may be less accurate for some of the more sophisticated bots that have become possible since the creation of more advanced technologies such as LLMs.
Indeed, certain bots can be difficult for even humans to distinguish~\cite{Cresci_2017}, and bot detection remains an ongoing problem.

Another limiting factor is the subjectivity of polarising topics.
Indeed, as touched upon at the end of Section~\ref{sec:signed_ego_network_analysis}, how polarising a topic is is not something that can be directly measured and it will be inconsistent across different contexts and cultures.
Unfortunately, subjectivity is an inevitable obstacle whenever one tries to analyse inherently qualitative phenomena such as natural language.
However, this is not the first time that topics' polarisation levels have been examined within the research areas of social media and sentiment analysis~\cite{Stieglitz_2013,Wakefield_2022,Weismueller_2023}.
What's more, numerous steps were taken to reduce subjectivity in the results as much as possible, given the number and size of the datasets employed in this work, it could reasonably be argued that the results represent accurate average measurements for each of the topics and communities as a whole.
For example, one of the terms from Table~\ref{tab:top_topics_user_negativities} labelled as Generic was ``pizza''. While it is possible to have a pizza-based conversation which is quite polarising (e.g. discussing pineapple as a pizza topic), ``pizza'' as a whole is not considered polarising as it is expected that the majority of interactions about pizza will be innocuous.

Additionally, when considering the results of the topic analysis (Subsection~\ref{sec:topic_analysis}), one must bear in mind that both the measures of negativities and the BERTopic model have inherent uncertainties that may become compounded when the models are used in tandem.
For the purposes of the current work, we have made the assumption that these effects are most likely to occur randomly and uniformly.
However, this is a simplifying assumption that should be tested in future research.

Lastly, it would be impossible to analyse all countries, topics and types of users, and the choice of which ones were included will influence how much can be gleaned from the results. However, as specific attention was made to collect data from a range of culturally diverse regions, as well as topics at different levels of polarity, it can reasonably be expected that these results are as accurate as feasibly possible.

\nolinenumbers

%
%
%
\bibliographystyle{plos2015}

\begin{thebibliography}{10}

\bibitem{Arnaboldi_2013}
Arnaboldi V, Conti M, Passarella A, Pezzoni F.
\newblock Ego networks in twitter: an experimental analysis.
\newblock In: Proceedings IEEE INFOCOM; 2013. p. 3459--3464.

\bibitem{Arnaboldi_2015}
Arnaboldi V, Passarella A, Conti M, Dunbar, RIM.
\newblock Online social networks: human cognitive constraints in Facebook and Twitter personal graphs
\newblock Elsevier. 2015

\bibitem{Barbieri_2022}
Barbieri F, Anke LE, Camacho-Collados J.
\newblock Xlm-t: Multilingual language models in twitter for sentiment analysis and beyond.
\newblock In: Proceedings of the Thirteenth Language Resources and Evaluation Conference; 2022. p. 258--266.

\bibitem{Centellegher_2017}
Centellegher S, López E, Saramäki J, Lepri B.
\newblock Personality traits and ego-network dynamics.
\newblock PLOS ONE;12(3). 2017

\bibitem{Conneau_2019}
Conneau A, Khandelwal K, Goyal N, Chaudhary V, Wenzek G, Guzm{\'a}n F, et~al.
\newblock Unsupervised cross-lingual representation learning at scale.
\newblock ICLR; 2019.

\bibitem{Cortes_1995}
Cortes C, Vapnik V.
\newblock Support-vector networks.
\newblock Machine learning. 1995;20:273--297.

\bibitem{Cresci_2017}
Cresci S, Di Pietro R, Petrocchi M, Spognardi A, Tesconi M.
\newblock The paradigm-shift of social spambots: Evidence, theories, and tools for the arms race
\newblock In: Proceedings of the 26th international conference on world wide web companion; 2017. p. 963--972

\bibitem{Dieng_2020}
Dieng AB, Ruiz FJ, Blei DM.
\newblock Topic modeling in embedding spaces.
\newblock Transactions of the Association for Computational Linguistics. 2020;8:439--453.

\bibitem{Dunbar_1992}
Dunbar RIM.
\newblock Neocortex size as a constraint on group size in primates.
\newblock Journal of human evolution. 1992;22(6):469--493.

\bibitem{Dunbar_1993}
Dunbar RIM.
\newblock Coevolution of neocortical size, group size and language in humans.
\newblock Behavioral and brain sciences. 1993;16(4):681--694.

\bibitem{Dunbar_1995}
Dunbar RIM, Spoors M.
\newblock Social networks, support cliques, and kinship.
\newblock Human nature. 1995;6:273--290.

\bibitem{Dunbar_1998}
Dunbar RIM.
\newblock The social brain hypothesis.
\newblock Evolutionary Anthropology: Issues, News, and Reviews: Issues, News, and Reviews. 1998;6(5):178--190.

\bibitem{Dunbar_2015}
Dunbar RIM, Arnaboldi V, Conti M, Passarella A.
\newblock The structure of online social networks mirrors those in the offline world.
\newblock Social networks. 2015;43:39--47.

\bibitem{Ester_1996}
Ester M, Kriegel, HP, Sander J, Xu, X.
\newblock A density-based algorithm for discovering clusters in large spatial databases with noise.
\newblock In Proceedings: KDD. 1996;96(34):226--231.

\bibitem{Fang_2015}
Fang A, Macdonald C, Ounis I, Habel P.
\newblock Topics in tweets: A user study of topic coherence metrics for Twitter data.
\newblock In Advances in Information Retrieval: 38th European Conference on IR Research (ECIR). 2016;38: pp.492--504.
  
\bibitem{Fukunaga_1975}
Fukunaga K, Hostetler L.
\newblock The estimation of the gradient of a density function, with applications in pattern recognition.
\newblock IEEE Transactions on information theory. 1975;21(1):32--40.

\bibitem{Gilbert_2009}
Gilbert E, Karahalios K.
\newblock Predicting tie strength with social media.
\newblock In: Proceedings of the SIGCHI conference on human factors in computing systems; 2009. p. 211--220.

\bibitem{Gottman_1993}
Gottman JM.
\newblock A theory of marital dissolution and stability.
\newblock Journal of Family Psychology. 1993;7(1). p.57--76.


\bibitem{Gottman_1999}
Gottman JM, Levenson RW.
\newblock What predicts change in marital interaction over time? A study of alternative models.
\newblock Family Process. 1999;38(2). p. 143---158.

\bibitem{Granovetter_1973}
Granovetter MS.
\newblock The strength of weak ties.
\newblock American journal of sociology. 1973;78(6):1360--1380.

\bibitem{Grootendorst_2022}
Grootendorst M.
\newblock BERTopic: Neural topic modeling with a class-based TF-IDF procedure.
\newblock arXiv preprint arXiv:220305794. 2022;.

\bibitem{Hart_1995}
Hart B, Risley TR.
\newblock Meaningful differences in the everyday experience of young American children.
\newblock Paul H Brookes Publishing; 1995.

\bibitem{Heydari_2024}
Heydari S, Iñiguez G, Kertész J, Saramäki J. 
\newblock Disentangling degree and tie strength heterogeneity in egocentric social networks. 
\newblock arXiv preprint arXiv:2403.19529; 2024.

\bibitem{Hill_2003}
Hill RA, Dunbar RI.
\newblock Social network size in humans.
\newblock Human nature. 2003;14(1):53--72.

\bibitem{Lancichinetti_2009}
Lancichinetti A, Fortunato S.
\newblock Benchmarks for testing community detection algorithms on directed and weighted graphs with overlapping communities.
\newblock Physical Review E—Statistical, Nonlinear, and Soft Matter Physics. 2009;80(1). 

\bibitem{Lau_2014}
Lau JH, Newman D, Baldwin T.
\newblock Machine reading tea leaves: Automatically evaluating topic coherence and topic model quality.
\newblock In: Proceedings of the 14th Conference of the European Chapter of the Association for Computational Linguistics; 2014. p. 530--539.

\bibitem{Lopez-Joya_2023}
Lopez-Joya S, Diaz-Garcia JA, Ruiz MD, Martin-Bautista MJ.
\newblock Bot Detection in Twitter: An Overview
\newblock International Conference on Flexible Query Answering Systems; 2023:131--144.

\bibitem{McAuley_2014}
Mcauley J, Leskovec J.
\newblock Discovering social circles in ego networks.
\newblock ACM Transactions on Knowledge Discovery from Data (TKDD); 2014:8(1), pp.1-28.

\bibitem{McInnes_2017}
McInnes L, Healy J, Astels S.
\newblock hdbscan: Hierarchical density based clustering.
\newblock In: Journal of Open Source Software, The Open Journal. 2017;2(11).

\bibitem{HDBSCAN_2024}
McInnes L, Healy J, Astels S
\newblock hdbscan 0.8.39. 2024 Oct 12 [cited 21 November 2024].
\newblock Available from: https://pypi.org/project/hdbscan/

\bibitem{Miritello_2013}
Miritello G, Moro E, Lara R, Mart{\'\i}nez-L{\'o}pez R, Belchamber J, Roberts SG, et~al.
\newblock Time as a limited resource: Communication strategy in mobile phone networks.
\newblock Social networks. 2013;35(1):89--95.

\bibitem{Perry_2018}
Perry BL, Pescosolido BA, Borgatti SP.
\newblock `Sociocentric and Egocentric Approaches to Networks' in Egocentric network analysis: Foundations, methods, and models
\newblock Cambridge university press. 2018;44.

\bibitem{Reimers_2019}
Reimers N, Gurevych I.
\newblock Sentence-bert: Sentence embeddings using siamese bert-networks.
\newblock arXiv preprint arXiv:190810084. 2019.

\bibitem{Reimers_2020}
Reimers N, Gurevych I.
\newblock Making monolingual sentence embeddings multilingual using knowledge distillation.
\newblock arXiv preprint arXiv:200409813. 2020;.

\bibitem{Roder_2015}
Röder M, Both A, Hinneburg A.
\newblock Exploring the space of topic coherence measures.
\newblock In Proceedings of the eighth ACM international conference on Web search and data mining. 2015; pp. 399-408.

\bibitem{Rosner_2014}
Rosner F, Hinneburg A, Röder M, Nettling M, Both A.
\newblock Evaluating topic coherence measures.
\newblock arXiv preprint arXiv:1403.6397. 2014.

\bibitem{Socievole_2012}
Socievole A, Marano S. 
\newblock Exploring user sociocentric and egocentric behaviors in online and detected social networks.
\newblock In 2nd IEEE Baltic Congress on Future Internet Communications (pp. 140-147). 2012.

\bibitem{Stieglitz_2013}
Stieglitz S, Dang-Xuan L.
\newblock Emotions and information diffusion in social media—sentiment of microblogs and sharing behavior.
\newblock Journal of management information systems. 2013;29(4):217--248.

\bibitem{Sunstein_1999}
Sunstein CR.
\newblock The law of group polarization.
\newblock Olin Law \& Economics Working Paper. 1999(91)

\bibitem{Sutcliffe_2015}
Sutcliffe AG, Wang D, Dunbar RIM.
\newblock Modelling the role of trust in social relationships.
\newblock ACM Transactions on Internet Technology (TOIT). 2015;15(4):1--24.

\bibitem{Tacchi_2022}
Tacchi J, Boldrini C, Passarella A, Conti M.
\newblock Signed ego network model and its application to Twitter.
\newblock IEEE BigData. 2022.

\bibitem{Tacchi_2023}
Tacchi J, Boldrini C, Passarella A, Conti M.
\newblock Cultural Differences in Signed Ego Networks on Twitter: An Investigatory Analysis.
\newblock In: Companion Proceedings of the ACM Web Conference; 2023. p. 1039--1049.

\bibitem{Tacchi_2024}
Tacchi J, Boldrini C, Passarella A, Conti M. Keep Your Friends Close, and Your Enemies Closer: Structural Properties of Negative Relationships on Twitter. 2024.
\newblock EPJ Data Science

\bibitem{Toprak_2022}
Toprak M, Boldrini C, Passarella A, Conti M.
\newblock Journalists’ ego networks in Twitter: Invariant and distinctive structural features.
\newblock Online Social Networks and Media. 2022;30:100207.

\bibitem{Wakefield_2022}
Wakefield RL, Wakefield K.
\newblock The antecedents and consequences of intergroup affective polarisation on social media.
\newblock Information Systems Journal.2023;33(3). p. 640--668.

\bibitem{Weismueller_2023}
Weismueller J,Gruner RL, Harrigan P, Coussement K, Wang S.
\newblock Information sharing and political polarisation on social media: The role of falsehood and partisanship.
\newblock Information Systems Journal. 2024;34(3). p. 854--893.

\bibitem{West_2020}
West BJ, Massari GF, Culbreth G, Failla R, Bologna M, Dunbar RIM, Grigolini P.
\newblock Relating size and functionality in human social networks through complexity. Proceedings of the National Academy of Sciences vol. 117, num. 31 p.18355--18358; 2020.

\bibitem{Zipf_1936}
Zipf, GK.
\newblock The psycho-biology of language: An introduction to dynamic philology. Routledge. 1936.

\end{thebibliography}

\end{document}